\documentclass{article}
\usepackage{LaThuileFPSpro}
\begin{document}
\title{ 
  THE TOTEM EXPERIMENT AT LHC
  }
\author{
  Giuseppe Latino \\
  (for the TOTEM Collaboration) \\
  {\em Siena University $\&$ Pisa INFN, Physics Dept., 
    Via Roma, 56 - 53100 Siena, Italy} \\
  {\em e-mail: giuseppe.latino@pi.infn.it}
  }
\maketitle

\baselineskip=11.6pt

\begin{abstract}
  The TOTEM experiment at the CERN LHC is here presented. 
  After an overview of the experimental apparatus, the measurement 
  of the total $pp$ cross section, elastic scattering and diffractive 
  phenomena is described. This physics programme will allow to 
  distinguish among different models of soft proton interactions.
\end{abstract}
\newpage
\section{Introduction}
The TOTEM experiment\cite{Totem_TDR} at the LHC is designed and optimized to measure 
the total $pp$ cross section with a precision of about 1$\div$2\,$\%$, to study the nuclear elastic 
$pp$ cross section over a wide range of the squared four-momentum transfer -t\footnote{
In the relativistic limit and for small scattering angles: $|t|\sim (p\theta)^2$, where $p$ is the 
proton momentum and $\theta$ is the scattering angle with respect to the original beam direction.
} 
($10^{-3}$\,GeV$^2$ $< |t| <$ 10\,GeV$^2$) and to perform a comprehensive physics programme on 
diffractive dissociation processes partially in cooperation with the CMS experiment.
In order to fulfill its physics programme, complementary to the programme of the general-purpose 
experiments at the LHC, the TOTEM experiment has to cope the challenge of 
triggering and recording events in the very forward region with a good acceptance for particles 
produced at very small angles with respect to the beam.
Based on the ``luminosity independent'' method the evaluation of the total cross section 
with such a small error will in particular require simultaneous measurement of the $pp$ elastic  
scattering cross section $d\sigma /dt$ down to $|t| \sim 10^{-3}$\,GeV$^2$ (to be extrapolated 
to $t$ = 0) as well as of the $pp$ inelastic interaction rate. 
In particular, the detection of elastically scattered protons at a location very close to the beam 
(indeed inside the beam-pipe itself) is required together with particle detection with the largest 
possible coverage in order to reduce losses on inelastic events detection to a few percent.
\begin{figure}[htb!]
  \vspace{4.0cm}
  \includegraphics{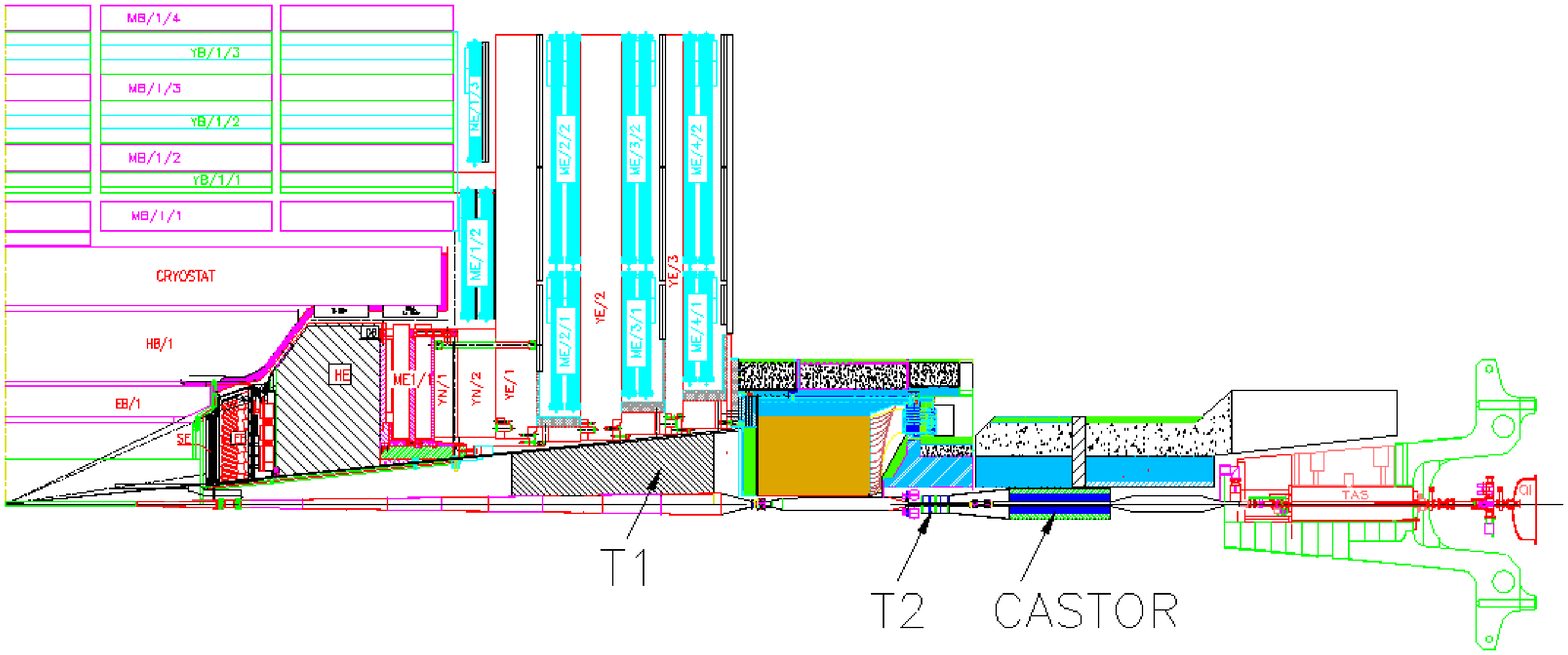}
  \vspace{3.0cm}
  \includegraphics{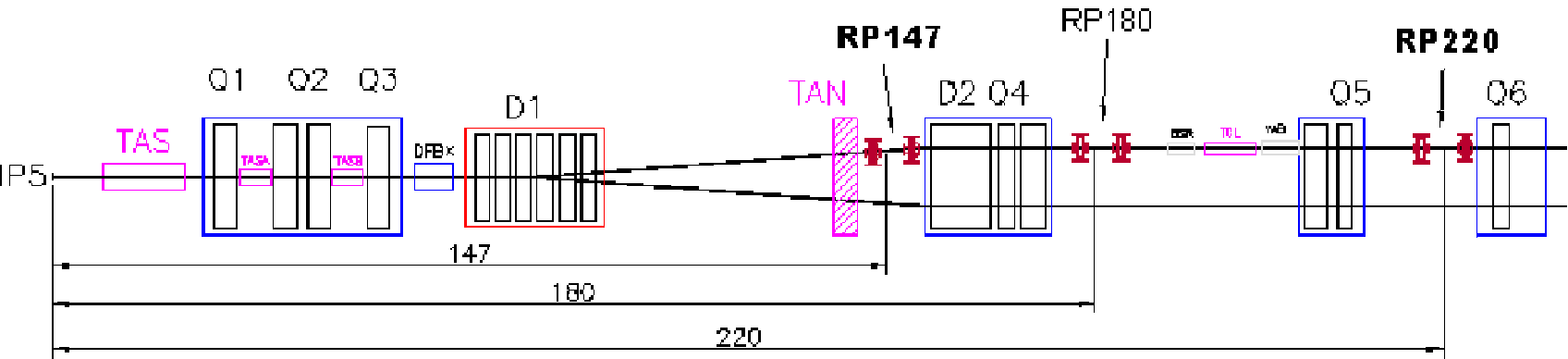}  
  \vspace{0.3cm}
  \caption{\it
    Top: the TOTEM forward trackers T1 and T2 embedded into the forward region 
    of the CMS detector. Bottom: TOTEM Roman Pots location along the LHC beam 
    line at a distance about 147 m (RP147) and 220 m (RP220) from the interaction point IP5,  
    RP180 being another possible location at the moment not equipped. 
    All TOTEM detector components are located on both sides of IP5.}
  \label{TOTEM_Det} 
\end{figure}

The TOTEM apparatus is located on both sides of the interaction point IP5, the same 
LHC experimental area as CMS\cite{CMS}. The T1 and T2 tracking detectors are embedded 
inside the forward region of CMS (see fig.\ref{TOTEM_Det}, top). 
Charged track reconstruction in the pseudo-rapidity\footnote{
 Pseudo-rapidity is defined as $\eta = -ln(tan\frac{\theta}{2})$.} 
range of $3.1 \le |\eta| \le 6.5$ is performed by these two inelastic telescopes 
and complemented at very high $|\eta|$ by detectors located in special movable beam-pipe insertion 
called ``Roman Pots'' (RP) which, being placed about 147 m and 220 m from IP5 (see fig.\ref{TOTEM_Det}, bottom), 
are designed to detect ``leading'' protons (scattered elastically or quasi-elastically from 
the interaction) at few mm from the beam center with a scattering angle down to few $\mu$rad.

The combination of the CMS and TOTEM experiments represents the largest acceptance detector ever 
built at a hadron collider which will also allow the study of a wide range of physics processes in 
diffractive interactions with an unprecedented coverage in rapidity. For this purpose the TOTEM data 
acquisition system (DAQ) is designed to be compatible with the CMS DAQ in order to make common data 
taking possible at a later stage.  

In the following, after a brief description of the experimental apparatus, the main 
features of the TOTEM physics programme will be presented.
%
%
%
%
\section{Detector Overview}
The TOTEM experimental setup comprises ``Roman Pots'' detectors to measure leading protons 
elastically scattered at very small angles within the beam pipe and 
the T1 and T2 inelastic telescopes providing charged track reconstruction for 3.1 $<$ $|\eta|$ $<$ 6.5 
with a 2$\pi$ coverage and with a very good efficiency in order to minimize losses (see fig.\ref{TOTEM_Det}). 
T1 and T2 track reconstruction will also allow trigger capability with acceptance grater 
than 95$\%$ for all inelastic events \footnote{ 
   About 99.5$\%$ of all non-diffractive minimum bias events and about 84 $\%$ of all diffractive 
   events have charged particles within the geometrical acceptance of T1 and T2 so that they are 
   triggerable with these detectors.}
as well as the reconstruction of the event interaction vertex so that 
background events (mainly from beam-gas interactions and halo muons)\cite{CMS_TOTEM_TDR} can be rejected.
Furthermore, the T1 and T2 detectors will provide tracking in front of the CMS HF (T1) and Castor (T2) very 
forward calorimeters so that the combination of these detectors can allow, for instance, a more complete study 
of ``rapidity gaps''\footnote{
   A rapidity gap, a region of pseudo-rapidity devoid of particles, is a typical signature for 
   diffractive processes which are characterized by a hadronic color singlet exchange with vacuum quantum numbers, 
   for which the Pomeron is one model.} 
and particle/energy flows in the very forward region. 

The read-out of all TOTEM sub-detectors is based on the digital VFAT chip\cite{VFAT} 
which is a tracking front-end ASIC specifically designed for the TOTEM experiment 
and characterized by trigger capabilities.
\subsection{Roman Pots}
\begin{figure}[htb!]
  \vspace{5.0cm}
  \includegraphics{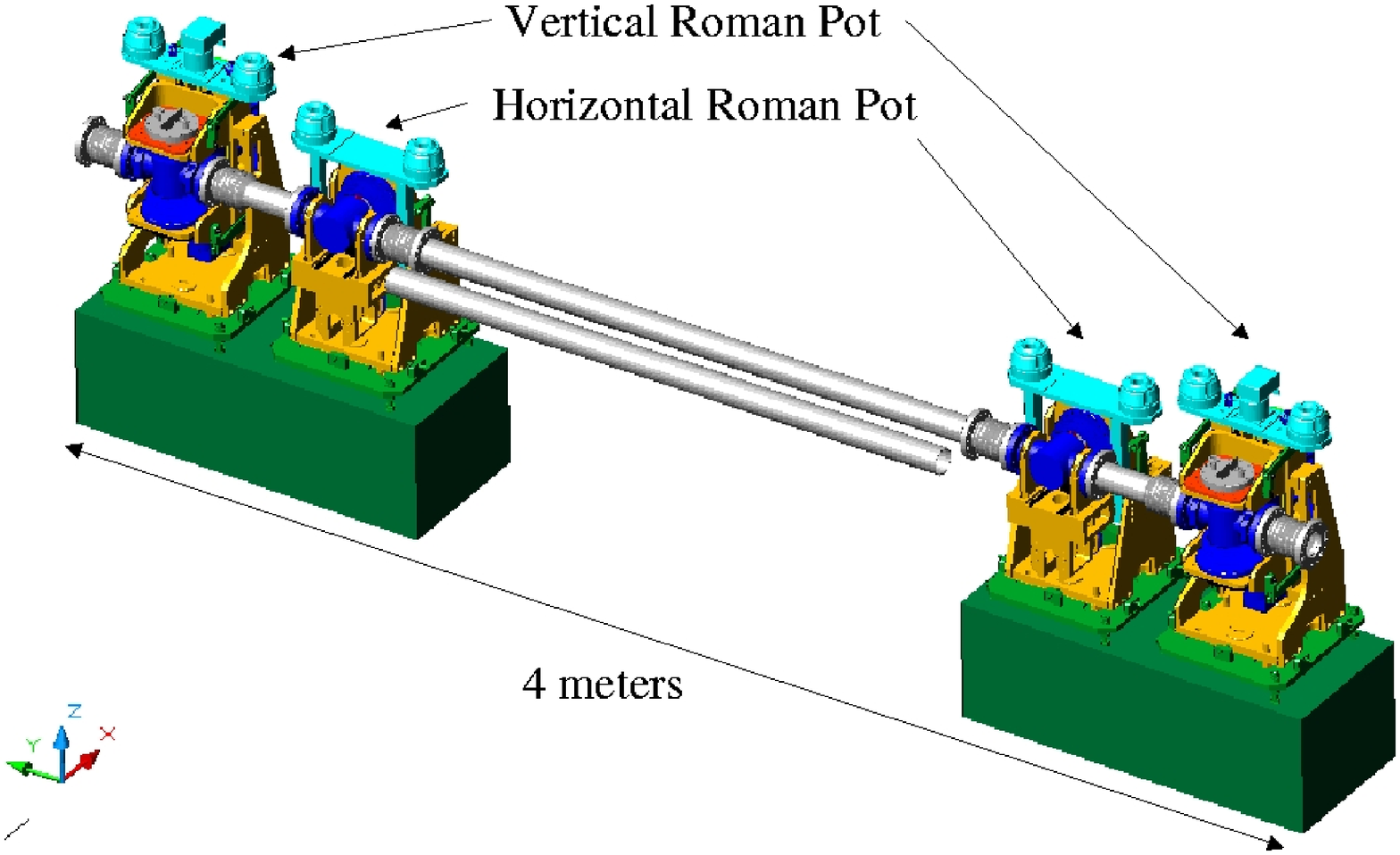}
   \hspace{7.5cm}
   \includegraphics{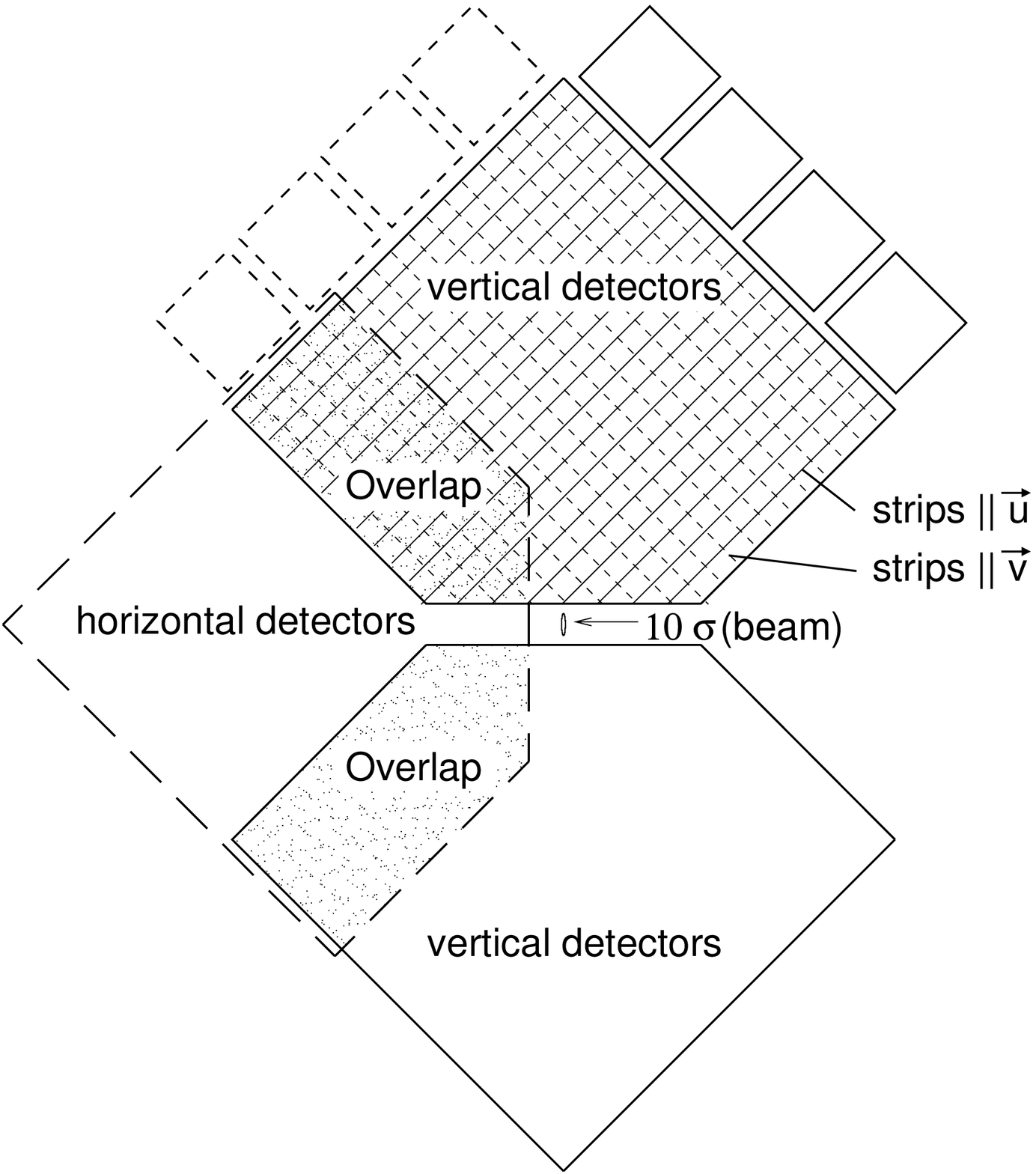}  
  \caption{\it
    Left: one TOTEM Roman Pot station.  
    Right: arrangement of silicon detectors inside two vertical and one horizontal
    pots at a RP unit.  
    \label{RP_Det} }
\end{figure}

The detection of very forward protons is performed by movable beam insertions, called ``Roman Pots'' 
(RP), hosting silicon detectors inside a secondary vacuum vessel (called ``Pot'') which are moved 
very close to the beam into the primary vacuum of the machine through vacuum belows. In this way 
the detectors can be put in a safe position when conditions of not stable beams are present 
(like at the very begin of a run), while are kept at the same time separated from the primary 
vacuum of the machine which is so preserved from an uncontrolled out-gassing of the detector materials.
Two RP stations are installed on both sides from the interaction point IP5 on the beam pipe 
of the outgoing beam at a distance of about 147\,m and at 220\,m, a position chosen according to 
the constraints given by the space available among the LHC machine components and by the special 
optics used by TOTEM.
A magnetic dipole between the two RP stations provides a magnetic spectrometer allowing an 
accurate proton momentum reconstruction. Each RP station is composed of two units 
(see fig.\ref{RP_Det}, left) in order to have a lever harm for local track reconstruction and trigger 
selections by track angle. 
Each unit consists of 3 pots, 2 approaching the beam vertically from the top and the bottom and one 
horizontally which completes the acceptance for diffractively scattered protons (see fig.\ref{RP_Det}, right).
The overlap of the detectors in the horizontal pots with the ones in the vertical pots allows a correlation 
of their positions via common particle tracks. This is used for the alignment of the three pots in an unit, 
the absolute alignment with respect to the beam being given by Beam Position Monitors (BPM) located in the 
vacuum chamber of the vertical pots.

Each pot contains a stack of 10 planes of novel silicon strip ``edgeless'' detectors, half of which 
have their strips oriented at an angle of $+45^o$ and half at an angle of $-45^o$ with respect to the 
edge facing the beam. Each plane has 512 strips with a pitch of 66 $\mu$m allowing a single hit
resolution of about 20 $\mu$m. In order to detect protons elastically scattered at angles down to few 
$\mu$rad at the RPs locations, these detectors need to have their active area edge moved as close 
to the beam as $\sim$ 1 mm. Consequently, their edge dead area had to be greatly minimized so that 
a new ``edgeless planar silicon'' detector technology has been developed for TOTEM RPs where a current 
terminating structure allows to reduce to only 50 $\mu$m the insensitive decoupling area 
 detector edge and sensitive volume\cite{RP_Silicon}.
For the same reason the stainless steel bottom foil of the pot (the one facing the beam) 
has been reduced to a thickness of 150 $\mu$m, while the pot window in front of the detector 
active area is 500 $\mu$m thick.

Irradiation studies on these silicon detectors, performed at the TRIGA reactor in Ljubljana at
different neutron fluxes up to $10^{14}~1\,\rm MeV\,n/cm^{2}$ and with 24\,GeV protons at CERN 
with a radiation up to $\rm 1.4\times 10^{14}\,p/cm^{2}$, have shown similar aging effects
as for devices using standard voltage terminating structures. 
Calculations of the diffractive proton flux hitting the detectors indicate 
that the present detectors will probably be working up to an integrated 
luminosity of about 1\,fb$^{-1}$. To cope with higher luminosities, TOTEM has 
initiated an INTAS project to develop radiation harder edgeless 
detectors~\cite{TOTEM_article}.

\subsection{T1 and T2 tracking detectors}
\begin{figure}[htb!]
  \vspace{5.5cm}
   \includegraphics{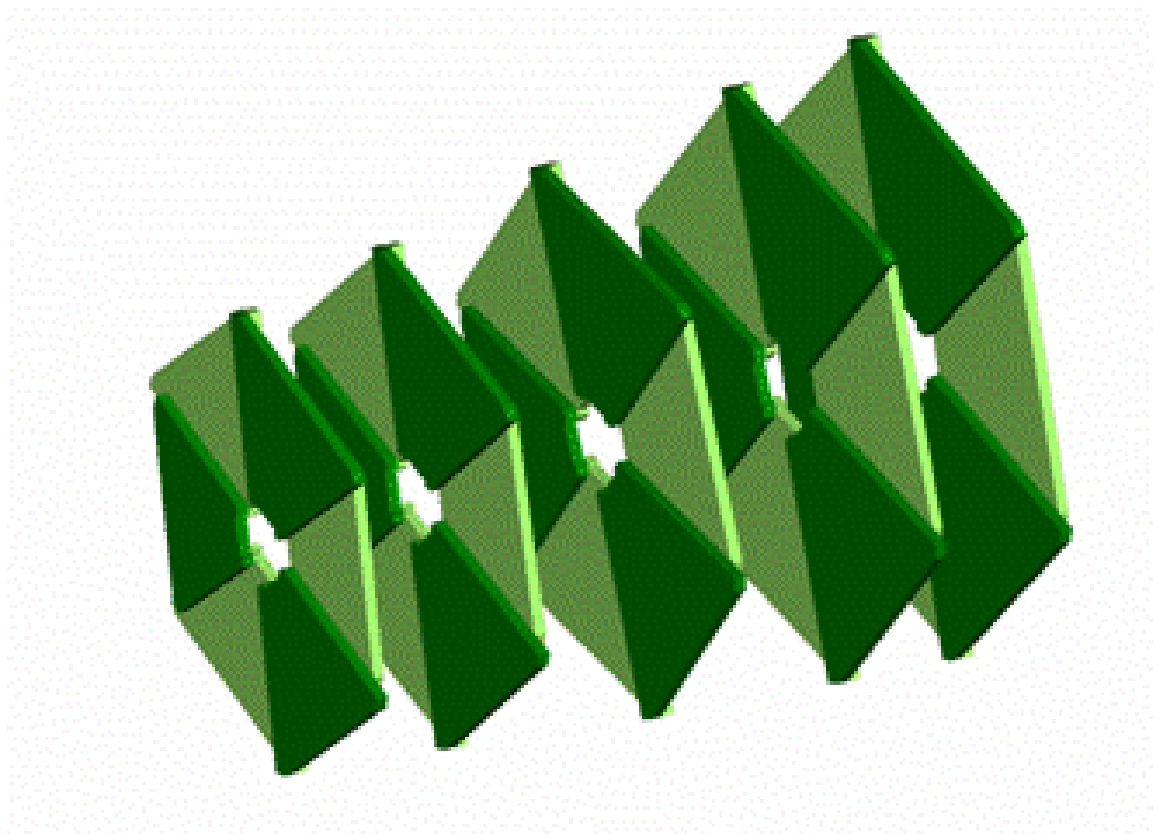}
  \hspace{7.cm}
   \includegraphics{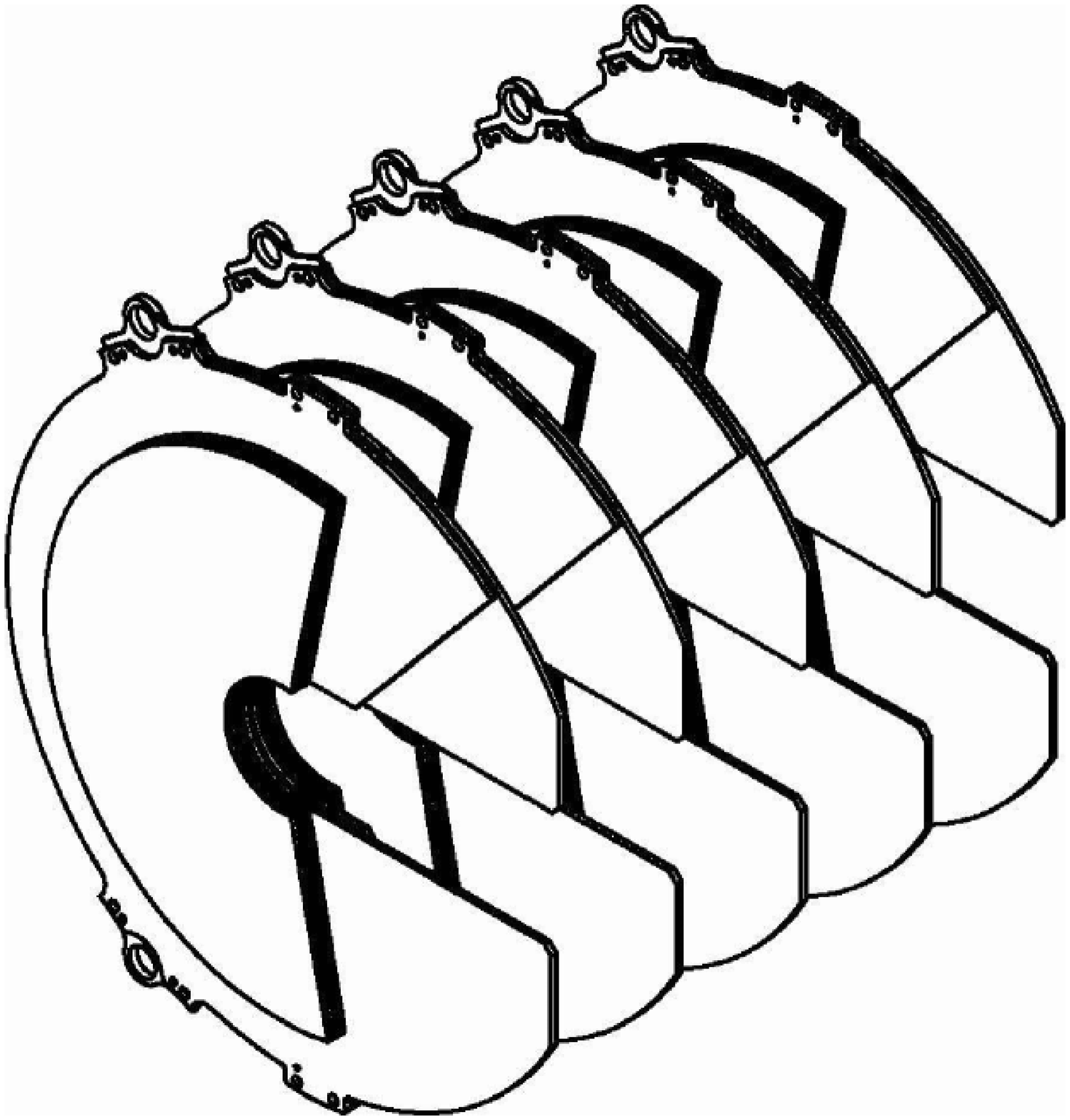}  
  \caption{\it
    Left: one arm of the TOTEM T1 telescope.  
    Right: one half-arm of the TOTEM T2 telescope.  
    \label{T1_T2_Det} }
\end{figure}

The T1 telescope covers the pseudo-rapidity range 3.1 $<$ $|\eta|$ $<$ 4.7 on both 
sides of IP5. Each telescope arm consists of five planes, equally spaced in $z$, formed 
by six trapezoidal ``Cathode Strip Chambers'' (CSC)\cite{Totem_TDR} 
(see fig.\ref{T1_T2_Det}, left). The detector sextants in each plane are rotated with respect 
to each other by angles varying from $-6^o$ to $+6^o$ in steps of $3^o$ in order to 
improve the pattern recognition for track reconstruction and 
to reduce the localized concentration of material in front of the CMS HF calorimeter. 
The TOTEM CSCs have a detector design similar to CMS CSC muon chambers with a gas gap 
of 10 mm and a gas mixture of Ar/CO$_2$/CF$_4$ ($40\%/50\%/10\%$). 
In these detectors the segmentation of cathode planes into parallel 
strips gives, combining their read-out with the one from anode wires, three measurements 
of the coordinates of the particle traversing the detector plane. 
Anode wires (with a pitch of 3 mm) give radial coordinate measurement which is also 
used for level-1 trigger information, while cathode strips (with a pitch of 5 mm) 
are rotated by $\pm$ $60^o$ with respect to the wires. Beam tests on final prototypes have 
shown a spatial resolution of about 0.8 mm when using VFAT digital read-out. 
Aging studies performed at the CERN Gamma Irradiation Facility have shown no loss of 
performance after an irradiation resulting in a total charge integrated on the anode wires of 
0.065 C/cm, which corresponds to an accumulated dose equivalent to about 5 years of running 
at luminosities of $10^{30}$cm$^{-2}$s$^{-1}$.  

The T2 telescope, based on novel ``Gas Electron Multiplier'' (GEM) technology\cite{GEM}, 
extends charged track reconstruction to the rapidity range 5.3 $<$ $|\eta|$ $<$ 6.5
\cite{Totem_TDR}. 
Placed 13.5\,m away from both sides of IP5, each T2 arm consists of a set 
of 20 triple-GEM detectors having an almost semicircular shape with 
an inner radius matching the beam pipe. Ten aligned detectors planes, 
mounted ``back-to-back'', are combined to form one T2 semi-arm on each side of 
the vacuum pipe (see fig.\ref{T1_T2_Det}, right). To avoid efficiency losses, the angular 
coverage of each detector is more than $180^o$.
GEMs are gas-filled detectors, already successfully adopted in other experiments such as 
COMPASS and LHCb, which have been considered for the design of the TOTEM very forward T2 telescopes 
thanks to their characteristics, in particular: good spatial resolution, excellent rate capability 
and good resistance to radiation. Furthermore, GEM detectors are also characterized by the advantageous 
decoupling of the charge amplification structure from the charge collection and read-out 
structure which allows an easy implementation of the design for a given apparatus.
The T2 GEMs\cite{T2_GEM} use the same baseline design as the ones adopted in COMPASS with 
a triple-GEM structure, realized by separating three GEM foils by thin (2\,mm) insulator spacers, 
adopted in order to reduce sparking probabilities while reaching typical total gas 
gains of about $8\times 10^3$ with a relatively low voltage (around 500\,V) applied to each 
GEM foil. The gas mixture is Ar/CO$_2$ (70$\%$/30$\%$). 
The read-out board, explicitly designed for TOTEM, has two separate layers with 
different patterns: one with 256x2 concentric circular strips, 80\,$\mu$m wide and with a pitch 
of 400\,$\mu$m, allowing track radial reconstruction, and the other with a matrix of 24x65 pads 
varying in size from 2x2\,mm$^2$ to 7x7\,mm$^2$ (for a constant 
$\Delta \eta \times \Delta \phi \sim 0.06 \times 0.017 \pi$) 
providing level-1 trigger information as well as track azimuthal reconstruction.
Final production detectors have been successfully tested in the 2007 beam test showing 
a spatial resolution in radial coordinate of about 100\,$\mu$m with digital VFAT read-out.
COMPASS triple-GEM detectors aging tests have shown that a charge up to 20 mC/mm$^2$ can be 
integrated on the read-out board without aging effects. This corresponds to run TOTEM for at least 
1 year at luminosities of $10^{33}$cm$^{-2}$s$^{-1}$. It is so assumed that TOTEM T2 triple-GEM can be 
operated during the first 3 years of LHC running.
\section{Physics Programme}
Given its unique coverage for charged particles at high rapidities,
TOTEM is an ideal detector for studying forward phenomena, including elastic 
and diffractive scattering.
Its main physics goals, precise measurements of the total $pp$ cross section 
$\sigma_{tot}$ and of the elastic scattering over a large range in $t$, 
are of primary importance in order to distinguish among 
different models of soft proton interactions. Furthermore, as energy flow and particle 
multiplicity of inelastic events peak in the forward region, the large rapidity 
coverage and the proton detection on both sides of the interaction point allow the study 
of a wide range of physics processes in inelastic and diffractive interactions. 
\subsection{Total $pp$ cross section}
\begin{figure}[htb!]
  \vspace{6.9cm}
    \includegraphics{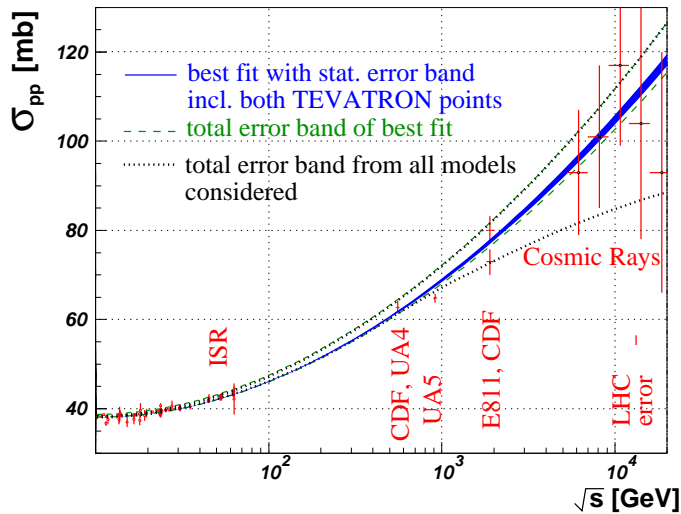}  
  \vspace{-0.2cm}
  \caption{\it
   Fits from the COMPETE collaboration to all available $pp$ and $p\bar{p}$ 
   scattering data~\cite{compete}.  
    \label{Tot_Xsec}}
\end{figure}

Fig.~\ref{Tot_Xsec} summarizes the existing measurements of 
$\sigma_{tot}$ from low energies up to collider and cosmic ray energies, 
also showing recent predictions for the energy dependence of $\sigma_{tot}$ 
by fitting all available $pp$ and $p\bar{p}$ scattering data 
according to different models~\cite{compete}. 
The dark error band shows the statistical errors to the best fit 
($\sigma_{tot} = 111.5 \pm 1.2 ^{+4.1}_{-2.1} $ mb for the LHC energy), the closest
dashed curves near it give the sum of statistical and systematic errors to the best fit
due to the discrepancy of the two Tevatron measurements, and the highest and lowest dotted 
curves show the total error bands (ranging in the 90$\div$130 mb interval) 
from all models considered. 
This large theoretical uncertainty is due to the current lack of a fully satisfactory
theoretical explanation of the cross section in low momentum transfer collisions, 
their description relying on phenomenological models to be tuned on existing data.
The large uncertainties of the cosmic ray data and the 2.6 standard deviations discrepancy 
between the two final results from the Tevatron give an extrapolation to 
the LHC energy ($\sqrt{s} = 14\,\rm TeV$) which is characterized 
by a wide range for the expected value of $\sigma_{tot}$, 
typically from 90 to 130\,mb, depending on the model used for the extrapolation.
More recent studies by other authors give predictions substantially within this range, with 
the exception of models with an explicit ``hard'' pomeron which give 
predictions at higher values~\cite{Khoze_Achilli}. 
TOTEM aims to measure $\sigma_{tot}$ with a precision down to $\sim$ 1\% (or $\sim$ 1 mb), 
therefore allowing to discriminate among the different models.

In absence of an accurate determination of the LHC luminosity
the measurement of $\sigma_{tot}$ will be based on the ``luminosity independent'' 
method which combines the optical theorem, relating $\sigma_{tot}$ to the imaginary 
part of the forward scattering amplitude and leading to the following equation:
\begin{equation}
 \mathcal{L}\, \sigma_{tot}^{2} = \frac{16 \pi}{1 + \rho^{2}} \cdot
 \left.\frac{dN_{el}}{dt} \right|_{t=0}
 \label{eqn_optical}
\end{equation}
with the total rate equation: 
\begin{equation}
 \mathcal{L} \,\sigma_{tot} = N_{el} + N_{inel}
 \label{eqn_totalrate}
\end{equation}
resulting in a system of 2 equations which can be solved for $\sigma_{tot}$ 
and $\mathcal{L}$, which are so expressed as a function of measurable rates:
\begin{equation}
 \sigma_{tot} = \frac{16 \pi}{1 + \rho^{2}} \cdot
 \frac{dN_{el}/dt |_{t=0}}{N_{el} + N_{inel}}
  \label{eqn_sigmatot}\\
\end{equation}
\begin{equation}
 \mathcal{L} = \frac{1 + \rho^{2}}{16 \pi} \cdot
 \frac{(N_{el} + N_{inel})^{2}}{dN_{el}/dt |_{t=0}} 
  \label{eqn_lumi}
\end{equation}
TOTEM will then measure $\sigma_{tot}$ and the luminosity $\mathcal{L}$ 
independently by experimentally measuring: the inelastic rate $N_{inel}$ consisting of
non-diffractive minimum bias events ($\sim$ 65\,mb at LHC) and
diffractive events ($\sim$ 18\,mb at LHC) 
which will be measured by T1 and T2; 
the total nuclear elastic rate $N_{el}$ ($\sim$ 30\,mb at LHC) and the nuclear part of the elastic 
cross section extrapolated to $t = 0$ (optical point) $dN_{el}/dt |_{t=0}$, 
measured by the Roman Pot system.
For the rate measurements it is important that all TOTEM detector systems have
trigger capability.
The expected uncertainty of the extrapolation to $t = 0$ depends on the acceptance 
for elastically scattered protons at small t-values, hence on the beam optics.
The $\rho$ parameter, defined by: 
\begin{equation}
\rho = \frac{\mathcal{R}[f_{el}(0)]}{\mathcal{I}[f_{el}(0)]}
\end{equation}
where $f_{el}(0)$ is the forward nuclear elastic scattering amplitude,
has to be taken from external theoretical 
predictions, e.g.~\cite{compete}. Since $\rho \sim 0.14$ enters only in a 
$1+\rho^2$ term, its impact is small.

A precise measurement of small scattering angles for the protons requires 
the beam angular divergence to be as small as possible, hence 
special runs with high machine optics $\beta^*$ are required. 
The consequent increase in beam size at the interaction point and the  
zero crossing angle technically related to this optics configuration also require 
a small number of bunches, in order to avoid extra interactions between the 
colliding beams inside the common vacuum chamber. 
Consequently the typical instantaneous luminosity for the 
TOTEM $\sigma_{tot}$ measurement at level of $\sim 1\%$, obtained with
an approved optics characterized by $\beta^* =$ 1540 m and 43 bunches, 
will be of the order of 10$^{28}$ cm$^{-2}$s$^{-1}$.
The requirement of a special injection optics for the optimal $\beta^* =$ 1540 m 
configuration makes it probably not available at the early beginning of LHC.
Another approved special beam optics with $\beta^* =$ 90 m (and a luminosity close to 
10$^{30}$ cm$^{-2}$s$^{-1}$), achievable without modifying the standard LHC injection 
optics, will allow a preliminary $\sigma_{tot}$ measurement at the level of 
about 5\% uncertainty as well as an excellent measurement of the momentum loss
of diffractive protons, opening the studies of soft and semi-hard
diffraction, the latter in combination with the CMS detectors.
After having understood the initial measurements and with improved beams at 
$\beta^{*} = 1540\,$m, a precision around 1\% should be achievable, provided 
an improved knowledge of the optical functions\footnote
{
The optical functions determine the explicit path of the particle through the magnetic elements 
and depend mainly on the position along the beam line (i.e. on all the magnetic elements
traversed before reaching that position and their setting which is optics 
dependent) but also on the particle parameters at the IP.}
and an alignment precision of the RP station better than 50 $\mu$m are obtained.

Given the high value of measured rates, the statistical error on $\sigma_{tot}$ measurement
will be substantially negligible after few hours of data taking even at low luminosity runs.
The vertex reconstruction will allow to largely reject the background
from beam-gas (dominant) and beam halo events to a negligible rate.
The systematic error for the measurement with $\beta^* =$ 90 m will be dominated by the extrapolation 
of nuclear elastic cross section to $t$ = 0 ($\sim$ 4\% for $|t|$ measured down to about $ |t| = 10^{-2}$ GeV$^2$), 
while for the $\beta^{*} = 1540\,$m measurement the total inelastic rate will give the main systematic 
uncertainty which will be dominated by trigger losses in Single Diffraction events ($\sim$ 0.8\%)\footnote
{
Dedicated studies have shown that Single and Double Diffraction events are
responsible for the major loss in the inelastic rate.
With a single-arm trigger (in coincidence with a leading proton in the 
opposite side RP for the single diffractive events) a fraction of these events, 
corresponding to $\sim$ 2.8\,mb, escapes detection. The lost events are mainly those 
with a very low mass (below $\sim$ 10\,GeV/c$^2$), since all their particles are 
produced at pseudo-rapidities beyond the T2 tracker acceptance.
The fraction of these lost events can be estimated by extrapolation to low masses
so to allow the determination of the total inelastic rate. 
For Single Diffraction, the extrapolated number of events differs from the simulation 
expectations by 4\%, corresponding to a 0.6\,mb uncertainty on the total
cross-section. The same estimate for Double Diffraction and Double Pomeron Exchange 
gives a 0.1\,mb and 0.2\,mb uncertainty, respectively.
}. 
The theoretical uncertainty related to the estimate of the $\rho$ parameter is expected to give 
a relative uncertainty contribution of less than 1.2\% (considering for instance the full error 
band on $\rho$ extrapolation as derived in ref~\cite{compete}). Combining all relevant 
uncertainties by error 
propagation for the equations~\ref{eqn_sigmatot} and~\ref{eqn_lumi}, also taking into account the 
correlations, gives a relative error of about 5\% (7\%) for the measurement of $\sigma_{tot}$ ($\mathcal{L}$) 
with $\beta^* =$ 90 m and of about 1$\div$2\% (2\%) for $\sigma_{tot}$ ($\mathcal{L}$) with 
$\beta^* =$ 1540 m~\cite{TOTEM_article}.
\subsection{Nuclear elastic $pp$ scattering}
\begin{figure}[htb!]
  \vspace{8.2cm}
   \includegraphics{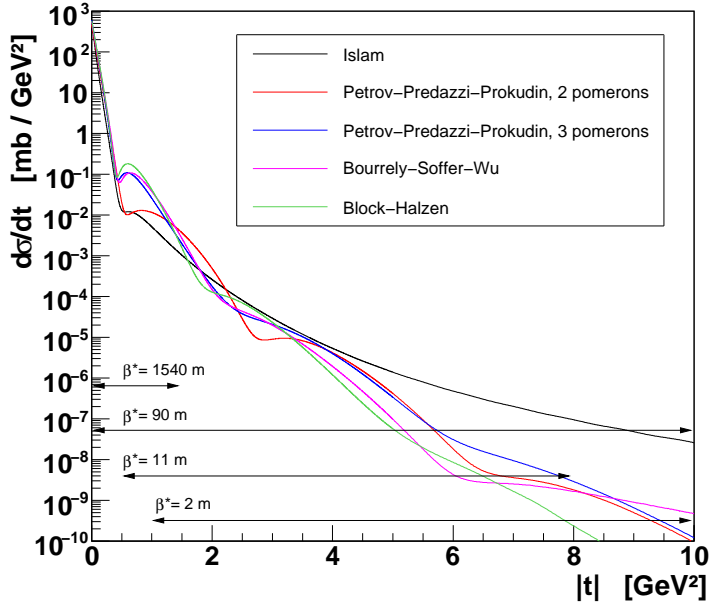}  
  \vspace{-0.2cm}
  \caption{\it  
    Differential cross section of elastic scattering at $\sqrt{s} = 14$\,TeV as predicted by 
    different models~\cite{elasticmodels}. The t-acceptance ranges for different optics settings 
    are also shown.   
    \label{Tot_El_Xsec} }
\end{figure}

Most of the interest in large impact parameter collisions is related to 
nuclear (hadronic) elastic scattering and to soft inelastic diffraction, both characterized,
e.g., by the exchange of hadronic colour singlets. 
Fig.~\ref{Tot_El_Xsec} shows the differential cross section of elastic $pp$ interactions 
at $\sqrt{s} = 14$\,TeV~\cite{kun1} as predicted by different models~\cite{elasticmodels}.
Several regions with different behavior can be distinguished when different 
$t$-ranges are considered at increasing $|t|$ (which means looking deeper into the proton 
at smaller distance).
The Coulomb region, where elastic scattering is dominated by one photon 
exchange ($d\sigma / dt \sim 1 / t^{2}$), is characterized by 
$|t| \le 10^{-5}\,{\rm GeV}^{2}$.
In the intermediate region for $|t|$ up to 0.002 GeV$^{2}$, 
the hadronic and Coulomb scattering 
interfere, complicating the extrapolation of the nuclear cross section to $t$ = 0.
The hadronic region, described in a simplified way, e.g.,  
by ``single-Pomeron exchange'' with an approximately exponential 
cross section ($d\sigma / dt \sim {\rm e}^{-B\,|t|}$) at its lower border, 
is expected for $0.002 < |t| < 0.4\,{\rm GeV}^{2}$.

The predictions of different models shown in Fig.~\ref{Tot_El_Xsec} have been obtained   
by fitting the differential cross section data at lower measured energies starting at the ISR energies. 
The shown results are based on the eikonal model.
The influence of the Coulomb scattering at lower $|t|$ values has been described with the help
of West and Yennie type of the total elastic scattering amplitude~\cite{kun1}. 
It is still an open question if a different approach could be used in order to remove the 
discrepancy on the elastic impact parameters introduced by the West and Yennie 
approach~\cite{kun1,kun2}.

It is evident that the interference region and the beginning of hadronic
region are important for the extrapolation of hadronic $d N_{el}/dt$ to $t=0$,
needed for determination of $\sigma_{tot}$.
The $t$-dependence of $B(t) = \frac{d}{dt} \ln \frac{d\sigma}{dt}$, 
shows slight model dependent deviations~\cite{kun1} from exponential shape, 
giving a theoretical uncertainty contribution to the systematic error of the total 
cross section measurement. The fit is typically performed with a 
quadratic polynomial parametrization in the $|t|_{min} < |t| < 0.25\,{\rm GeV}^{2}$ interval, 
where $|t|_{min}$ depends on the acceptance for protons elastically scattered at small angles, 
which is related to the beam angular divergence. 
The expected uncertainty on the extrapolation to $t$ = 0 will be related to $|t|_{min}$ 
($|t|_{min}$ $\sim$ $0.002\, (0.04)\,{\rm GeV}^{2}$ for $\beta^{*} = 1540\, (90)\,$m), hence 
it will depend on the beam optics.
The diffractive structure of the proton is then expected in the $|t| > 0.4$ GeV$^2$ region.
Finally, for $|t| \geq 1.5 \div 3$ GeV$^2$ there is the domain of central elastic collisions 
at high $|t|$ which might be described by perturbative QCD, e.g., in terms of three gluon exchange 
with a predicted cross section proportional to $|t|^{-8}$~\cite{donn}. 

We can see from fig.~\ref{Tot_El_Xsec} that there is a model dependence of the predictions 
which is very pronounced at high $|t|$.
To discriminate among different models it is thus important to precisely
measure the elastic scattering over the largest possible $t$-region.
As shown in fig.~\ref{Tot_El_Xsec} TOTEM can study different $t$-ranges depending on 
the LHC optics setting. Under different beam optics and running conditions  
TOTEM will cover the $|t|$-range from $2 \times 10^{-3}\,$GeV$^{2}$ to about 10\,GeV$^{2}$ 
spanning the elastic cross section measurement for over 11 orders of magnitude.
\subsection{Diffraction and inelastic processes}
\begin{figure}[htb!]
  \vspace{5.0cm}
   \includegraphics{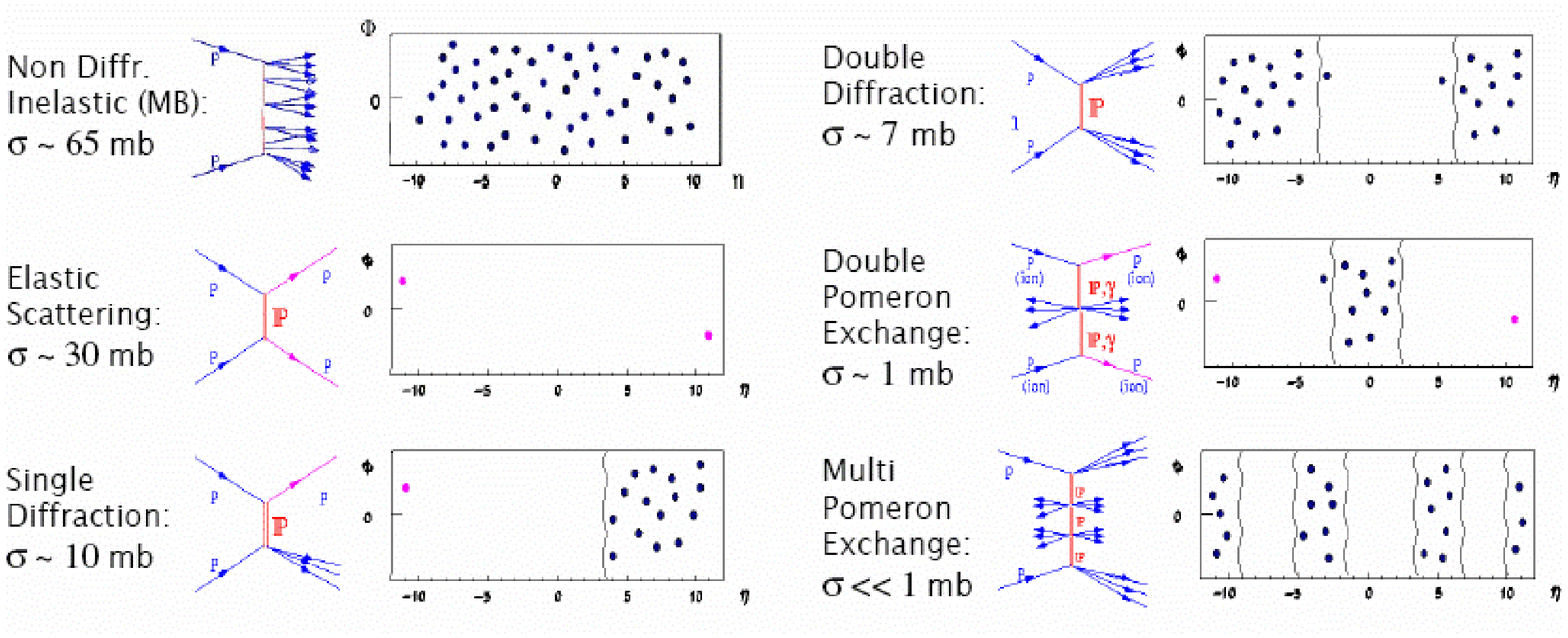}  
  \caption{\it  
	    Typical event topology for non diffractive (Minimum Bias) and diffractive processes 
    in the pseudorapidity-azimuth plane. 
    The associated cross sections, as estimated for the LHC, are also reported.   
    \label{Evt_Topology} }
\end{figure}

Fig.~\ref{Evt_Topology} shows the typical event topology for 
non diffractive (Minimum Bias) and diffractive processes together with the associated 
cross sections, as expected at the LHC.  
Diffractive scattering processes (Single Diffraction, Double Diffraction, 
``Double Pomeron Exchange'', and higher order ``Multi Pomeron'' processes) 
together with the elastic scattering one, represent about 50\,\% of the 
total cross section. Nevertheless, many details of these processes with close 
ties to proton structure and low-energy QCD are still poorly understood. 
The majority of diffractive events exhibits intact (``leading'') protons in 
the final state, characterized by their $t$ and by their fractional momentum loss
$\xi \equiv \Delta p/p$, most of which (depending on the beam optics) can be 
detected in the RP detectors.
Already at an early stage, TOTEM will be able to measure $\xi$-, $t$- and 
mass-distributions in soft Double Pomeron and Single Diffractive events.
The integration of TOTEM with the CMS detector will offer the possibility
of more detailed studies of the full structure of diffractive events, with the optimal 
reconstruction of one or more sizeable rapidity gaps in the particle distributions which 
can be obtained when the detectors of CMS and TOTEM will be combined for common data taking 
with an unprecedented rapidity coverage, as detailed in ref~\cite{CMS_TOTEM_TDR}. 
For this purpose the TOTEM triggers, combining information from the inelastic 
detectors and the silicon detectors in the RPs, are designed to be also incorporated 
into the general CMS trigger scheme. 

TOTEM will also provide a significant contribution to the understanding of very 
high energy cosmic ray physics as it will give accurate informations on the basic properties
of $pp$ collisions at the maximum accelerator energy. A challenging
issue in astrophysics is in fact represented by primary cosmic rays in the PeV (10$^{15}$ eV) 
energy range and above. The LHC center of mass energy corresponds to a 100 PeV energy for 
a fixed target collision in the air. At the same time the LHC will provide a very high event 
rate relative to the very low rate of cosmic ray particles in this energy domain.
Several high energy hadronic interaction models are nowadays available describing  
the nuclear interaction of primary cosmic ray entering the upper atmosphere and 
generating air showers. They predict energy flow, multiplicity and other quantities 
of such showers which characteristics are related to the nature of the primary interaction 
and to the energy and composition of the incident particle.
There are large differences among the predictions of currently available
models, with significant inconsistencies in the forward region. Several quantities can be 
measured by TOTEM and CMS and compared with model predictions, among which: 
energy flow, elastic/total cross section, fraction of diffractive events, 
particle multiplicity. The study of the features of diffractive and inelastic events as 
measured in TOTEM and CMS may thus be used to validate/tune these generators~\cite{CMS_TOTEM_TDR}.
\section{Summary and Conclusions}
The TOTEM experiment will be ready for data taking since the very beginning of the LHC 
start. Running under all beam conditions, it will be able to perform an important and 
exciting physics programme involving total and nuclear elastic scattering $pp$ cross 
section measurements as well as diffractive precesses studies. 
Special high ${\beta}^*$ runs will be needed in order to perform an optimal measurement 
of total $pp$ cross section at the level of $\sim 1\div 2$ $\%$ (for ${\beta}^*$ = 1540 m).
An early measurement is foreseen with ${\beta}^*$ = 90 m (more easily achievable) with a 
relative error at the level of $\sim$ 5 $\%$. The measurement of elastic scattering in the range 
$10^{-3}\,{\rm GeV}^{2} < |t| < 10\,{\rm GeV}^{2}$ will allow to distinguish among a wide 
range of predictions according to current theoretical models. 
Finally, a common physics programme with CMS on soft and hard diffraction as well as 
on forward particle production studies will also be pursued.
\section{Acknowledgements}
I'm very grateful to the Conference Organizers, Giorgio Bellettini, Giorgio Chiarelli 
and Mario Greco for their kind invitation and warm hospitality at this very profitable 
and pleasant annual appointment for the HEP Community in La Thuile. I would like also 
to thank my Colleagues of the TOTEM Collaboration for their hard work in the development 
of the experiment and for precious input to this presentation.

\end{document}